\documentclass[aps,pre,twocolumn,showpacs]{revtex4}

\usepackage{amsmath}

\usepackage{graphicx}

\begin{document}

\title{Attraction of Spiral Waves by Localized Inhomogeneities with
Small-World Connections in Excitable Media}
\author{Xiaonan Wang, Ying Lu, Minxi Jiang, Qi Ouyang \cite{atr}}

\address
{Department of physics, Peking University, Beijing 100871, P.R.
China}

\begin{abstract}
Trapping and un-trapping of spiral tips in a two-dimensional
homogeneous excitable medium with local small-world connections is
studied by numerical simulation. In a homogeneous medium which can
be simulated with a lattice of regular neighborhood connections,
the spiral wave is in the meandering regime. When changing the
topology of a small region from regular connections to small-world
connections, the tip of a spiral waves is attracted by the
small-world region, where the average path length declines with
the introduction of long distant connections. The "trapped"
phenomenon also occurs in regular lattices where the diffusion
coefficient of the small region is increased. The above results
can be explained by the eikonal equation and the relation between
core radius and diffusion coefficient.
\end{abstract}


\maketitle

\section{Introduction}

Spiral waves are characteristic structures of excitable media that
have been observed in many extended systems such as
reaction-diffusion media \cite{r1,zhou1,zhou2}, aggregating
colonies of slime mold \cite{lee}, and heart tissues \cite{r2},
where they are suspected to play an essential role in cardiac
arrhythmia and fibrillation. Sudden cardiac death resulting from
ventricular fibrillation is generated from the fragmenting or
breakup of spiral waves \cite{r3,r3-1,Leo}. Spiral waves are prone
to a variety of instabilities \cite{ou1,ou2,ou3}, one of which is
meander instability \cite{ou4,bar1}, where spiral tips follow a
hypocycloid trajectory instead of moving around a small circle.
Due to Doppler effect, this spiral may undergo a transition from
ordered spiral patterns to a state of defect-mediated turbulence
\cite{ou3}. In the meandering regime, the spiral tips can be made
to drift and controlled by external influences \cite{r4} or
localized inhomogeneities of defects \cite{r5}.

After the concept of small-world connections was proposed by Watts
and Strogatz \cite{r10}, it has quickly attracted much attention
because this kind of connections exists commonly in real world,
such as in social systems \cite{r6}, neural networks \cite{r7} and
epidemic problems \cite{r8}. Different studies show that a little
change of the network connections can essentially change the
features of a given medium, and plays a very important role in
determining the dynamic behavior of a system.

In numerical simulations, a spatially extended system can be
approximately regarded as a network consisting of a number of
sites connected with certain topology. In this way localized
inhomogeneities can be achieved by changing the topology of the
network. In heart tissues, pacemakers dominate the dynamics of the
travelling wave behavior and control the heart rhythm. We
hypothesize that this happens because the characters and the
structures of the local cells are different from other cardiac
muscle cells. Could a small-world network describe one of the
characters of the pacemaker? To answer this question, we changed
the widely used regular network in spiral wave study to a
small-world network in part of the system to investigate its
effects.

In the next section, we study the effect of a local small-world
network on the behavior of spiral waves. We show that this special
region is a dynamic attractor for spiral tips. In section III, we
compare the effect of the small-world network with that of
changing diffusion coefficient in the local region, and show that
they are equivalent. We give a discussion and conclude our study
in the last section.

\section{The effect of small-world network}

The model we used is the two-variable FitzHugh-Nagumo model
\cite{r9} with local nearest-neighbor couplings in a region of
$N_{1}\times N_{2}$:
\begin{eqnarray}
{\frac{du_{i,j}}{dt}}
&{=}&{(a-u_{i,j})(u_{i,j}-1)u_{i,j}-v_{i,j}+D_u
\nabla ^{2}u_{i,j}}  \label{eq:a1} \\
{\frac{dv_{i,j}}{dt}} &{=}&{\varepsilon
(bu_{i,j}-v_{i,j})+D_v\nabla ^{2}v_{i,j}}  \label{eq:a2}
\end{eqnarray}
where $i=1,2,...,N_{1}$, $j=1,2,...N_{2}$; $u_{i,j}(t)$ and
$v_{i,j}(t)$ is dimensionless excitable variable and recovery
variable respectively; $D_u$ and $D_v$ are diffusion coefficients
of the two variables. The Laplacien in the last term can be
approximated as:
\begin{eqnarray}
\nabla ^{2}u_{i,j} \cong \frac{1}{h^{2}}%
(u_{i-1,j}+u_{i+1,j}+u_{i,j-1}+u_{i,j+1}-4u_{i,j})  \label{eq:b1} \\
\nabla ^{2}v_{i,j} \cong \frac{1}{h^{2}}%
(v_{i-1,j}+v_{i+1,j}+v_{i,j-1}+v_{i,j+1}-4v_{i,j}).  \label{eq:b2}
\end{eqnarray}
When $0<a<1$, $b \geq 0$, $D_v << D_u$, $\epsilon << 1$, the
equation describes an excitable medium, which can be regarded as a
simplified model for cardiac tissues. In the following discussion,
we set the control parameters as follows: $N_{1}=N_{2}=256$,
$a=0.1$, $b=1.0$, $\epsilon =0.005$, $D_u=0.33$, $D_v=0$. No-flux
boundary condition is applied in the simulation.

Using vertical gradient distribution initial condition, we first
create spiral waves in regular lattices (Fig. 1(a)). In this case,
the spiral tip follows a hypocycloid trajectory, showing a typical
sign of meandering state \cite{ou4}, see Fig. 1(c). We then create
a small-world network in a small local region $\Omega $ of a size
$N_{s}\cdot N_{s}(N_{s}<<N_{1})$, where the tip of spiral waves
locates. The small-world network is created in the following way:
With the probability $p_{s}(0\leq p_{s}\leq 1)$, we reconnect
every edge in the region $\Omega $ from one of its original vertex
to another vertex chosen randomly in the region \cite{r10}(see
Fig. 1(b)). The change of connections leads to the change of
"diffusion" mode. Supposing node[i][j] is connected with
node[$x_1$][$y_{1}$], node[$x_{2}$][$y_{2}$], \ldots node
[$x_{k}$][$y_{k}$], then the "diffusion" term in the Eqs (1) (2)
becomes:
\begin{eqnarray}
\nabla ^{2}u_{i,j}\cong \frac{1}{h^{2}}%
(u_{x_1,y_1}+u_{x_2,y_2}+...+u_{x_k,y_k}-k\cdot u_{i,j})
\label{eq:d1} \\
\nabla ^{2}v_{i,j}\cong \frac{1}{h^{2}}%
(v_{x_1,y_1}+v_{x_2,y_2}+...+v_{x_k,y_k}-k\cdot v_{i,j})
\label{eq:d2}
\end{eqnarray}

\begin{figure}[tbp]
\centering
\includegraphics[width=8cm]{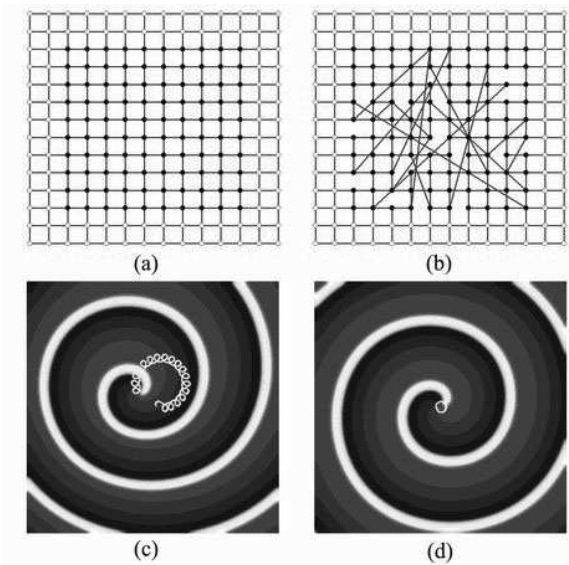}
\caption{Spiral pattern in different media. (a) a regular network
$p_{s}=0$; (b) a small-world network, $p_{s}=0.1$; (c) spiral wave
and its tip's motion in the regular network. the white curve is
the trajectory of the spiral tip; (d) spiral wave and its tip's
motion in the small-world network. $N_s=10$.} \label{FIG.1}
\end{figure}

Introducing the small-world network region in the reaction medium
greatly influences the motion of the spiral tip. We find that the
small-world network region can attract the spiral tip as it passes
through the region. After that, the spiral tip rotates around its
boundary, as shown in Fig. 1(d). Because the topological structure
of the local small-world network is generated randomly under the
above mentioned rule, the attraction only occurs under a certain
parameter range with certain probability. To characterize the
attraction property of the small-world network, we use the
attraction probability $p_{a}$ as an order parameter, which can be
obtained by repeating (50 times in our work) the simulation using
the same control parameters but with different small-world network
connections. Our simulation results show that the most influential
factor to $p_{a}$ is the small-world creation probability $p_{s}$.
As show in Fig. 2(a), $p_{a}$ increases with the increase of the
small-world $p_{s}$. When $p_{s}=0$ (corresponding to a regular
network), the spiral tip cannot be attracted; when $p_{s}=1$
(corresponding to a random network), the tip can be attracted with
probability 1. Between $0<p_{s}<1$, there is a transition where
the attraction probability $p_{a}$ increases rapidly. The
transition point $p_c$ can be defined as the value of $p_s$ when
$p_a=0.5$.

\begin{figure}[tbp]
\centering
\includegraphics[width=6cm]{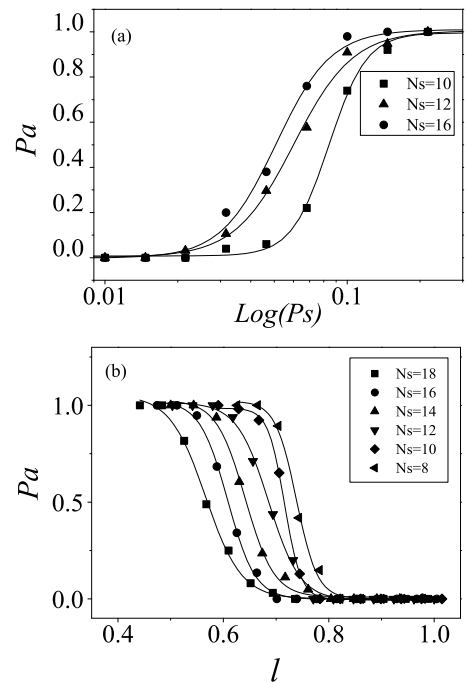}
\caption{(a) transition curves of $p_{a}$ as a function of $p_{s}$
with different $N_{s}$. The line is the sigmoidal fitting
$p_{a}=1-1/(1+\exp ((p_s-p_{sc})/dp_s))$; (b) transition curves of
$p_{a}$ as a function of $l$ with different $N_{s}$. The line is
the fitting of $p_{a}=1/(1+\exp ((l-l_{c})/dl)).$ Here $l_{c}$ is
the critical length where $p_{a}=0.5$, and $dl$ is the transition
width.} \label{FIG.2}
\end{figure}

One of the most important characters of small-world network is the
reducing of average path length while keeping the clustering
coefficient almost constant \cite{r10}. Define the normalized
average path length $l$ of small-world network as $l=L/L_{0}$,
where $L$ is the average path length of small-world network in
$\Omega $ \cite{r11} and $L_{0}$ is the average path length of
regular network in $\Omega $. $l$ will decrease from $1$ to
$\varepsilon (\varepsilon>0)$ when $p_{s}$ changes from 0 to 1.
From numerical simulations, we find the same type of transition
curve of $p_a$ as a function of $l$, as shown in Fig. 2(b),
indicating that the major effect of the small-world topology to
the behavior of spiral waves is the decrease of average path
length. In addition, defining the critical length $l_c$ as the
value of $l$ when $p_a=0.5$, we find that the critical length
decreases linearly as the increase of the size of $\Omega $, see
Fig. 3(a).

\section{Compare with increasing the diffusion coefficient}

Our simulation results suggest that the major effect of the
small-world network on spiral tip movement comes from the long
distance connections, which lead to shortening the average path
length ($l$) and increasing the diffusion speed. If the above
suggestion is correct, the phenomenon of spiral tip attraction
should also occur when we locally increase the diffusion constant
in Eqs. (1) and (2). In this part of work, we increase the
diffusion coefficient in a small circular region $\Omega $ by $D $
times and keep the system with regular connections. In the
following discussion, we will use $D_{u}^{\Omega}$ to denote the
diffusion coefficient in region $\Omega$, and use $D_{u}^{0}$ for
the region outside of $\Omega$, so that $D_{u}^{\Omega}=D\cdot
D_{u}^{0}$. We find that the spiral tip can be attracted and
travel around the boundary of region $\Omega$ when $D$ is large
enough and $R>R_0$.($R$ is the radius of $\Omega $, $R_0$ is the
core radius of spiral when $D_u^{\Omega}=D_u^0$). At a given $R$,
we can define two values $D_1$ and $D_2$: A temporal attraction
occurs when $D_1<D<D_2$. In that case, the tip can be trapped for
a short period and then escapes; the "trapped" time increases with
$D$. When $D>D_2$, the tip can be trapped for a long enough
period. Define $D_c$ as the mean value of $D_1$ and $D_2$, the
plot of $D_c$ with different $R$ is shown in the Fig. 4.

The above simulation results suggest that the increase of
diffusion speed in the small region is responsible for the
attraction of the spiral tip. To quantitatively compare the two
systems, we analyze the diffusion terms of the two systems. In the
small-world network, because of the long distance connections, the
average distance between nodes declines as $p_{s}$ increases from
$0$ to $1$. In a network model of a reaction-diffusion system,
this effect can be in a sense translated from the decrease of the
average path length between nodes while keeping the distance of
two neighboring nodes $h$ constant, to the decrease of the step
length $h$ while keeping the network regular. The normalized
average path length $l$ can also describe the relative change of
$h$. From this argument, the diffusion items of the small-world
network can be expressed as:
\begin{eqnarray}
D_{u}^{0}\frac{1}{(lh)^{2}}
(u_{i-1,j}+u_{i+1,j}+u_{i,j-1}+u_{i,j+1}-4u_{i,j})  \label{eq:f1} \\
D_{v}^{0}\frac{1}{(lh)^{2}}
(v_{i-1,j}+v_{i+1,j}+v_{i,j-1}+v_{i,j+1}-4v_{i,j})  \label{eq:f2}
\end{eqnarray}

At the critical point, for a fixed $R$, the diffusion terms in two
systems should be the same. So that $D_{u}^{\Omega}\cdot
1/h^{2}=D_c\cdot D_{u}^{0}\cdot 1/h^{2}=D_{u}^{0}\cdot 1/(l_c\cdot
h)^{2}$, which gives $D_c\cdot l_c^{2}=1$ . As presented in Fig.
3(b), our simulation fits this analysis within the range of error.
This result indicates that our proposition is reasonable. The
attracting effect of the small-world network comes from the
decline of $l$ inside the inhomogeneous area $\Omega $.

\begin{figure}[tbp]
\centering
\includegraphics[width=6cm]{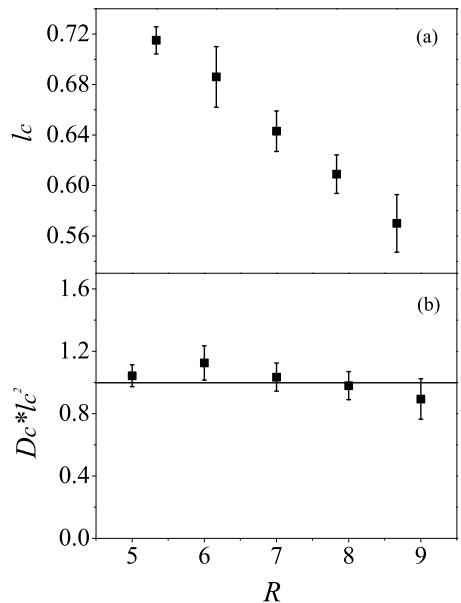}
\caption{(a) The critical length $l_c$ as a function of the
region's radius $R$, $(R=Ns/2)$, the error bar is $dl$,(see the
caption of Fig. 2(b) about the fitting); (b) $ D_c\cdot l_c^{2}$
as a function of $R$.} \label{FIG.3}
\end{figure}

\section{Discussion}

A question should be answered before fully understanding the
effect of the small-world network in the dynamics of spiral tips:
what is the mechanism for the spiral tip attraction? In the
following discussion, we give an explanation with the eikonal
equation, Luther equation \cite{Luther}, and the relation between
diffusion coefficient and of spiral core radius. According to the
analysis of the spiral tip dynamics given by Hakim and Karma
\cite{r13}, for the steady rotational movement of a spiral tip in
an excitable medium, the core radius $R$ as a function of
diffusion coefficient can be written as:
\begin{equation}
R=\frac{D_{u}}{c_{0}}(\frac{b\cdot
K}{B_{c}-\frac{2D_{u}}{W}})^{3/2} \label{eq:g1}
\end{equation}
Where $D_{u}$ is the diffusion coefficient of activator ($u $);
$c_{0}$ is the speed of plane wave; $b$, $K$ and $B_{c}$ are all
constants. $W$ is the constant width of the excited region. In the
simulation, We assume that at the boundary of $\Omega$ exists a
"virtual" gradient region of $D_u$ which links the outside and
inside regions. For a given $R$ of the region $\Omega
(R>R_{0}=R(D_{u}^{0}))$, the "trapped" motion of the spiral tip
requires a specific value of $D_{u}$, satisfying equation
(\ref{eq:g1}). When $D_{u}<D_{u}^{\Omega}$, the spiral tip will
enter the gradient region where the system can find the required
$D_{u}$, so that the spiral tip will rotate around the gradient
area at the boundary of $\Omega $; on the other hand, when
$D_{u}>D_{u}^{\Omega}$, the "trapped" motion can not be sustained
by the central region. From this argument, at critical point, we
will have $D_{u}(R) =D_{u}^{\Omega}=D_{c}\cdot D_{u}^{0}$. As
shown in Fig. 4, Our simulation results are consistent with this
analysis with the range of error.

\begin{figure}[tbp]
\centering
\includegraphics[width=6cm]{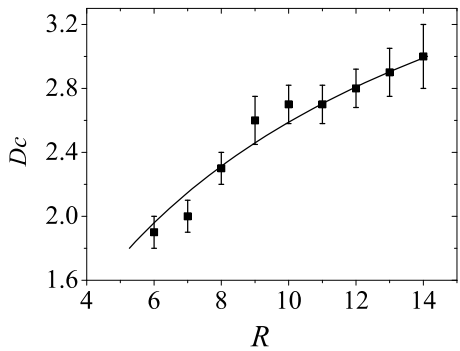}
\caption{The critical diffusion coefficient $D_c$ as a function of
$R$, where the line is the best fitting with Equation
(\ref{eq:g1}). The error bars are estimated using $D_1$ and $D_2$.
} \label{FIG.4}
\end{figure}

To prove the "trapped" state of tip motion is a stable state, we
apply the eikonal equation, which determines the relation between
the curvature of a travelling wave front and its speed in an
excitable medium, and the Luther relation, which describes the
relation between the speed of chemical waves and the diffusion
coefficient of activator \cite{Luther}. the eikonal equation is:
$N=C-D_{u}\cdot \kappa $, where $N$ is the normal wave speed,
$\kappa $ is the local curvature of the wave front; the Luther
equation is: $c=\alpha \cdot \sqrt{D_{u}},$ where $\alpha $ is a
constant. Insert the Luther equation into eikonal relation we
have:
\begin{equation}
N=\alpha \cdot \sqrt{D_{u}}-D_{u}\cdot \kappa  \label{eq:h1}
\end{equation}
Taking partial derivative of R in Eq. (\ref{eq:h1}), we get:
\begin{equation}
\frac{\partial N}{\partial R}=[\frac{1}{2}\alpha
(D_{u})^{-1/2}-\kappa ]\cdot \frac{\partial D_{u}}{\partial R}
\label{eq:h2}
\end{equation}
At the spiral tip we have $N=0$, so that:
\begin{equation}
\frac{\partial N}{\partial R}\left\vert _{tip}\right.
=-\frac{1}{2}\kappa _{tip}\cdot \frac{\partial D_{u}}{\partial R}
\label{eq:h3}
\end{equation}

In our system, assuming there exists a continuous change of
$D_{u}$ at the boundary of region ${\Omega}$, we have
$\frac{\partial Du}{\partial R}<0$. Thus the equation
(\ref{eq:h3}) indicates that $\frac{\partial N}{\partial
R}\left\vert _{tip}\right  >0$. That means, if we introduce a
small deviation from the "trapped" motion of the spiral tip, the
system will return to the "trapped" state spontaneously, because
we have $N<0$ inside the region $\Omega$, and $N>0$ outside the
region $\Omega$.($\hat{n}$ points to the center of ${\Omega}$
region).

In conclusion, we find that in an excitable system local change of
topological structure can trap the spiral tip. This ability comes
from the increase of diffusion speed. We prove this by increasing
the diffusion coefficient. We also give a theoretical explanation
using the eikonal equation, Luther equation, and the relation
between core radius and diffusion coefficient, which fits well
with the results of simulation. We should note that there are
other situations where the tip of spiral waves can be trapped in a
given area. For example, L\'{a}z\'{a}r {\it et al.} reported that
self-sustained chemical waves can rotate around a central obstacle
in an annular 2-D excitable system, and the wave fronts in the
case of an annular excitable region are purely involutes of the
central obstacle in the asymptotic state \cite{Zolly}. Obviously
this phenomenon is beyond our analysis. So that more work should
be done to fully understand the attract effect of local
inhomogeneities in an excitable reaction-diffusion system.

\begin{acknowledgments}
This work is partly supported by the grants from Chinese Natural Science
Foundation, Department of Science of Technology in China and Chun-Tsung
Scholarship at Peking University.
\end{acknowledgments}

\end{document}